# Predict; Don't React for Enabling Efficient Fine-Grain DVFS in GPUs

Srikant Bharadwaj, Shomit Das[‡], Kaushik Mazumdar, Bradford Beckmann, Stephen Kosonocky[*]

*Advanced Micro Devices, Inc.*

## ABSTRACT

With the continuous improvement of on-chip integrated voltage regulators (IVRs) and fast, adaptive frequency control, dynamic voltage-frequency scaling (DVFS) transition times have shrunk from the microsecond to the nanosecond regime, providing additional opportunities to improve energy efficiency. The key to unlocking the continued improvement in V/f circuit technology is the creation of new, smarter DVFS mechanisms that better adapt to rapid fluctuations in workload demand.

It is particularly important to optimize fine-grain DVFS mechanisms for graphics processing units (GPUs) as the chips become ever more important workhorses in the datacenter. However, GPU's massive amount of thread-level parallelism makes it uniquely difficult to determine the optimal V/f state at run-time. Existing solutions—mostly designed for single-threaded CPUs and longer time scales—fail to consider the seemingly chaotic, highly varying nature of GPU workloads at short time scales.

This paper proposes a novel prediction mechanism, PCSTALL, that is tailored for emerging DVFS capabilities in GPUs and achieves near-optimal energy efficiency. Using the insights from our fine-grained workload analysis, we propose a wavefront-level program counter (PC) based DVFS mechanism that improves program behavior prediction accuracy by 32% on average for a wide set of GPU applications at 1μs DVFS time epochs. Compared to the current state-of-art, our PC-based technique achieves 19% average improvement when optimized for Energy-Delay$^2$ Product (ED$^2$P) at 50μs time epochs, reaching 32% power efficiencies when operated with 1μs DVFS technologies.

## 1 INTRODUCTION

Dynamic Voltage Frequency Scaling (DVFS) techniques continue to improve the energy efficiency of modern computing architectures generation after generation [1]. The expansive benefits of DVFS come from the cubic relationship of voltage to power consumption owing to the basic dynamic power equation $P = CV^2Af$, where frequency $f$ also reduces with voltage $V$. The insight behind DVFS is that systems exhibit phased behavior where their performance has varying dependence on per-component operating frequencies, which in modern systems can be modified dynamically with supply voltage. Due to this phased behavior, dynamically adjusting frequencies and voltages can minimize unnecessary power consumption, resulting in more power-efficient architectures when managed properly.

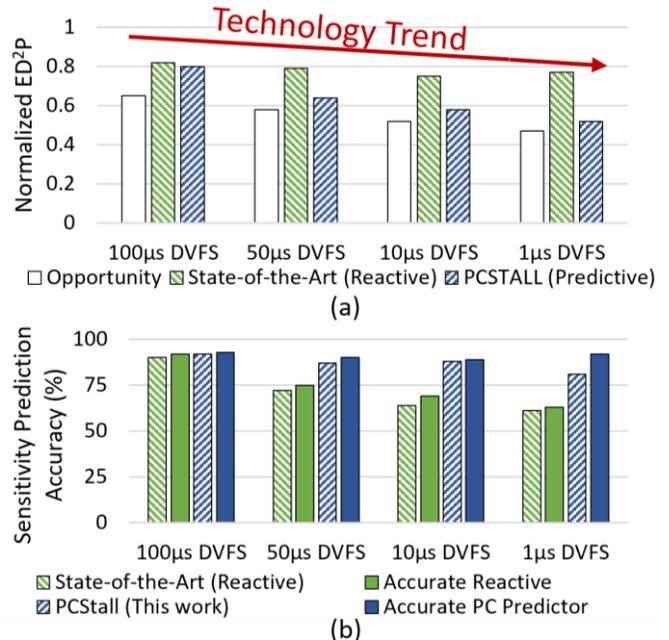

Figure 1: (a) Opportunity for improving GPU ED$^2$P at different DVFS time epochs. (b) Program behavior prediction accuracy of the state-of-the-art sensitivity estimation model [20] compared to an even more accurate reactive (theoretical) estimation model and the predictive mechanism proposed in this paper.

The recent advances in the design of on-chip voltage regulators [2, 3, 4, 5, 6] open new opportunities for finer frequency control and fundamentally push the power efficiency of future systems to higher limits [2]. Unlike earlier off-chip voltage regulator designs, modern circuit technology enables the integration of regulators within the chip. These integrated voltage regulators (IVR) were introduced just a few years ago [6] and continue to improve, providing ever-decreasing response latencies [7]. In particular, digital low dropout (LDO) regulators and switched regulators have emerged as an efficient new class of IVRs because of their low area footprints and small quiescent current requirements [8, 9, 10, 11]. Thus, they are increasingly being adopted in modern multicore DVFS systems [12, 13, 14, 15].

Complementary to IVR improvements, clock generation units have also undergone significant recent advances. IVRs can be combined with Phase-Locked Loops (PLLs) and digital frequency synthesizer (DFS) solutions where integer/fractional dividers are used to generate more traditional coarse-grain power/clock (V/f) domains. However, an alternative emerging clocking scheme gaining popularity is voltage adaptive Frequency-Locked Loops (FLLs) [6, 8, 7]. With these FLLs, V/f domains can be created by providing a clock frequency proportionate to the domain's supply voltage which in turn is controlled by IVRs. When supply voltage is adjusted, these new FLLs facilitate frequency adjustment transition times within few nanoseconds [6, 16, 17, 18, 8, 7] primarily due to their shorter settling times, allowing for iso-frequency time epochs of only a few microseconds or less. As a result, these faster IVRs can extract more energy efficiency and thereby justify the integration of more, finer-grain DVFS-enabled V/f domains within a system [2, 19, 7].

One particularly promising, yet uniquely challenging, avenue for fine-grain DVFS is its application to Graphics Processing Units (GPUs) which are increasingly being used for general compute purposes and machine intelligence applications. Compared to CPUs, GPUs operate over wider dynamic voltage ranges [20, 21, 22] and thus, have a higher potential for power savings, which has been demonstrated in GPU system prototypes [12, 23]. However, GPUs support 10,000 or more active threads, which is multiple orders of magnitude greater than CPUs. As a result, GPUs have far more chaotic behavior, which is far harder to predict than previously evaluated CPU scenarios. In addition, while DVFS techniques for CPUs have been extensively explored over the previous decade [1, 10, 24, 25], research on applying DVFS to GPUs has only started taking off in recent years [20, 26, 27, 28, 12, 29].

The majority of these previous works, in both the CPU and GPU space, focused on adopting optimal operating frequencies at time periods or epochs of *milliseconds* to *hundreds of microseconds*. However, nanosecond DVFS transition times now allow for moving to microsecond-level time epochs where we observe additional energy efficiency opportunity. Figure 1(a) uses a simulation of a 64-compute-unit GPU (see Section 5) running a wide variety of applications (see TABLE II) and shows the near-maximum opportunity for dynamically optimizing efficiency at different DVFS time epochs compared to static operation. As shown, DVFS technology at 1μs time epochs can reduce Energy-Delay$^2$ Product (ED$^2$P) by nearly 30% more versus DVFS at hundreds of milliseconds. Meanwhile, most previously proposed DVFS techniques [12, 20, 26, 27, 28] have not been evaluated for these shorter times epochs.

Maximizing the efficiency improvement with fine-grain DVFS requires an accurate *estimation* of frequencies' impact on program behavior of elapsed time epochs and *prediction* of future time epochs. Most of the prior DVFS techniques use analytical [24, 25] or machine learning estimation models [4, 30, 31] to determine a heuristic relationship between performance and the optimal operating frequency or phase. These models [1], designed for single-threaded CPUs, rely on *estimating* the execution time of any given workload or work segment at different frequencies before selecting an optimal frequency. Such techniques have also been extended to GPU Compute Units (CUs) [20], as shown in Figure 2(a) but lead to lower accuracies. Beyond the problem of estimation, existing *reactive* mechanisms use the estimation of the current epoch to set the frequencies of subsequent time epochs. However,

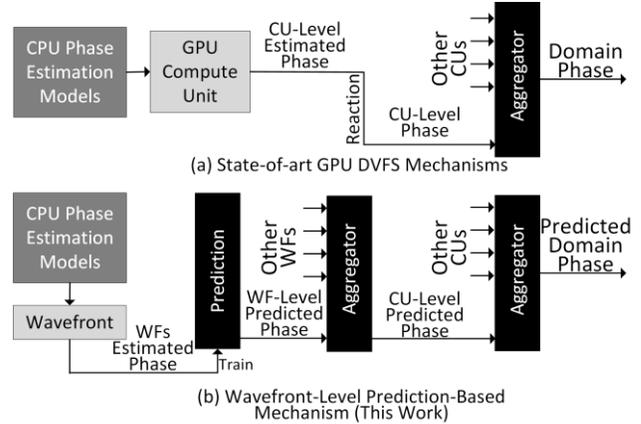

**Figure 2: (a) State-of-art GPU techniques [20] modify previous CPU models and apply it to each Compute Unit (CU). (b) This work proposes applying simpler models to wavefronts (warps) in combination with a predictor for improving the phase prediction accuracy.**

such reactive mechanisms result in low accuracy as shown in Figure 1(b).

Several previous works [10, 20, 24, 32, 33] have attempted to improve the accuracy of these sensitivity estimates, but even with the most accurate estimates in a loss of at least 16% loss of potential efficiency when used *reactively* at 50μs. Furthermore, the inability of reactive policies to achieve maximum power savings amplifies as we approach finer-grain DVFS time epochs, as seen in Figure 1(a). Thus, leveraging the improvement in circuit voltage and frequency control technologies requires an understanding of workloads at these fine-grain time epochs as well as the development of a new mechanism which can provide better *prediction* accuracy than a traditional reactive system.

To that end, we analyze GPU workloads at these fine-grain time scales and observe that GPUs showcase varying phase behavior, thus reducing the predictability of the system. For this, we adopt a methodology that measures a workload's frequency sensitivity at fine-grain time epochs. Specifically, we measure instructions executed over fixed-time epochs as a proxy for work done over the epochs (Section 3.1). We then define and employ a frequency *sensitivity* metric (Section 3.2) to identify the fine-grain phase patterns. We observe that GPUs show highly varying phase behavior (Section 3.3) during kernel execution. Further, we identify that the massive multithreading inherent to GPUs prevent the direct application of CPU-based models for determining phase behavior at fine-grain time scales (Section 4.1). These observations showed that existing simple reactive methods result in a high loss of power efficiency because of varying GPU workload behavior.

Based on these observations, we propose a novel prediction mechanism (as opposed to reaction), PCSTALL, that predicts the frequency sensitivity of future fine-grain time epochs in GPUs resulting in near optimal energy efficiency. PCSTALL (Section 4.4) utilizes the Program Counter (PC) value of wavefronts (or warps) to predict the phase and delivers higher sensitivity prediction accuracy (up to 81%), as shown in Figure 2(b). Consequently, our approach provides 19% average improvement in the ED$^2$P compared to state-of-art techniques at 50μs time epochs and reaches up to 32% improvement with 1μs DVFS time epoch technology.

## 2 RELATED WORK AND BACKGROUND

DVFS is a widely used technique for improving the energy efficiency. The general idea of DVFS is to dynamically change the voltage and frequency of operation to match the current requirements of a processing element (or group of processing elements). While the classical objective of DVFS has been to improve power savings [1], recent works have also proposed DVFS for improving performance under a power constraint [25, 34, 35]. Exploiting DVFS requires a mechanism that can estimate performance at different frequencies and then adjusts the clock frequency and voltage supply to optimize an objective function. Several previous works have proposed a wide variety of mechanisms for estimating workload phase behavior and assumed several different voltage regulator technologies. The following subsections summarize these works.

### 2.1 Technology Improvements in Fast-Fine-Grain DVFS

The voltage regulator transition latency significantly influences the overall efficiency achieved by a DVFS-enabled system. Thus, we first summarize the history of the voltage regulator technology.

**Challenges in Traditional DVFS.** Traditional V/f scaling using off-chip regulators exhibited voltage transition latencies between V/f states in the order of hundreds of microseconds [36]. This allowed for DVFS management to be performed by a firmware [37, 38, 39]. In addition, traditional PLLs have long latency re-lock times on the order of multiple microseconds which creates a restrictive upper bound on efficient DVFS times for off-chip regulation. In contrast, the higher speed DVFS solutions targeted in this paper require a much lower latency frequency adjustment mechanism.

**Integrated Voltage Regulators.** With the advent of different types of integrated voltage regulators (IVRs) [2, 3, 4, 40, 41, 6], it has now become possible to significantly reduce transition times between voltage-frequency states. This is mainly because they can provide a fast transient response to dynamically varying loads. There are multiple families of IVRs based on the methodology used for voltage regulation: linear resistive LDOs, switched capacitor regulators, and inductor-based buck converters.

Several efforts have been made to improve the design of these IVRs allowing transition times in the order of few nanoseconds. Switched regulators were traditionally known to be expensive and complicated in terms of die area. However, with the advancement of packaging technologies, such as using substrate layers for air core inductors [6] or embedding magnetic arrays inside package [42], die area is no longer a large concern and switched regulators have since been shown to provide fast responses [16, 6, 43]. LDOs on the other hand, operate on feedback-based resistive voltage division and thus can be implemented with entirely digital components. Recent works have shown LDOs provide fast settling times in the range of nanoseconds [18, 8, 17] and they have been proven to be beneficial in commercial designs [13, 12, 14, 23].

**Phase-Locked Loops.** Alternatively, high-speed digital frequency synthesis (DFS) techniques, developed for adaptive clock stretching [13, 7], can also be used to quickly select a derivative frequency relative to a locked maximum reference frequency generated from a PLL. Recent commercial products [13] have been shown to have a clocking system that can stretch a reference clock by a programmable amount in the nanosecond time frame. Several studies [2, 44, 19] have used the technology to show that fine-grain DVFS mechanisms can improve the overall energy efficiency of a system.

Earlier DVFS works from a decade ago proposed DVFS policies in the order of hundreds of milliseconds [25], which reduced to a single millisecond [45] policies a few years later. Recent works [20] on GPU DVFS techniques have described time epochs as low as tens of microseconds. While technology improvements have made commercial GPU products and prototypes showcase frequency transition times within nanoseconds [12, 15, 23], there has been minimal work done in supporting the reduced DVFS time epochs. The trend of technological improvement in IVRs, PLLs, and FLLs promises DVFS iso-frequency time epoch duration times to reduce from the current status of hundreds of microseconds to a few microseconds in the near future in commercial products.

Apart from the temporal aspect, on-chip regulators also impact the spatial effectiveness of a DVFS-based system by enabling multiple V/f domains within a chip [7]. With multi-core CPUs becoming commonplace, there has been a lot of CPU-related work managing multiple V/f domains [46, 47]. Several studies have concluded that managing multiple separate domains provides improved energy efficiency as compared to a single domain [34, 47, 48, 49]. Thus, our work focuses on fast as well as multiple V/f domain support in GPUs. In the remainder of this paper, we use the word *fine-grain* to describe these targeted fast-transitioning, finely tunable domains.

### 2.2 DVFS Control Mechanisms

There has been extensive literature [24, 10, 32, 33, 50, 51] on controlling DVFS systems for both CPUs and GPUs [20, 26, 27, 28]. DVFS mechanisms have also been extended to work for memory subsystem and CPU + Memory combined management [45, 31, 52]. Such control policies require monitoring dynamic system behavior, estimating performance at other possible V/f states, and adjusting the operating V/f states for future time epochs to meet power-performance targets. Thus, exploiting the DVFS mechanism to its maximum potential requires the ability to predict the future performance of a system at different operating V/f states. If we know the performance at each operating state, we can transition to an optimal frequency given an objective function. The general approach towards this challenge is to adopt heuristic methods to predict the performance of a future time epoch at different frequencies. Solutions usually involve two key challenges. First, when executing a work segment at one operating frequency for a time epoch, it is difficult to *estimate* the performance at other frequencies for the same work segment. This is a major challenge because workloads often demonstrate non-linear performance behavior at different frequencies. Second, even with accurate estimation of an elapsed time epoch, it is a challenge to *predict* what the workload behavior will be for a future time epoch, especially for workloads with highly varying behavior. Prior solutions to these two challenges are described next.

## 2.3 Estimating Frequency Sensitivity

Estimating frequency sensitivity to performance of an elapsed work segment has been studied in-depth by several previous works. While earlier works studied the estimation of frequency sensitivity using a linear scaling approach [53], performance counters [24, 10, 32, 33] have been shown to be more accurate. These approaches mainly fall into two major categories: (a) sampling, and (b) analytical. The sampling models estimate a scaling factor (sensitivity) by executing a workload at two different V/f points. The general idea is that the compute performance scales with frequency, while asynchronous memory phases remain constant. Most importantly, the intuition behind this generalization is that any CPU workloads (even multithreaded) can be approximated by a single in-order thread of execution. Although CPUs can execute out-of-order, memory stalls can still dominate portions of execution which makes this approximation acceptable [10]. Overall, such models estimate the time (T) to execute a workload as follows:

$$T_{f_2} = T_{async} + \frac{f_1}{f_2} T_{core\,@f1}$$

These models allow one to estimate the execution time at any frequency $f_2$ after executing at a frequency $f_1$. Separating the execution time into core and asynchronous-memory slices is a challenge and has been studied in depth [20, 24, 10, 25, 32]. Many previous works have proposed improving execution time estimation by considering a variety of system behaviors such as memory-level-parallelism [10], store stalls [20], and multi-core workloads [46, 54]. We briefly discuss some of the key mechanisms previously proposed.

**Stall Model.** This model estimates the asynchronous time spent by the processor by measuring the time it spent stalled for memory responses [24]. It ignores memory level parallelism and assumes that no computation is done during a memory operation.

**Leading Load.** This model [24, 32, 33] was proposed to incorporate memory level parallelism. A leading load is defined as a memory load operation performed when there are no other memory loads already in flight. The model accounts for the asynchronous time by measuring the latency of leading loads.

**Critical Path.** The Critical Path [10] model was proposed to consider a realistic memory subsystem and DRAM. The model book-keeps the timestamps such that loads on the critical path of execution are taken into consideration. The latencies of these critical path loads are then combined for estimating the asynchronous time spent by the core.

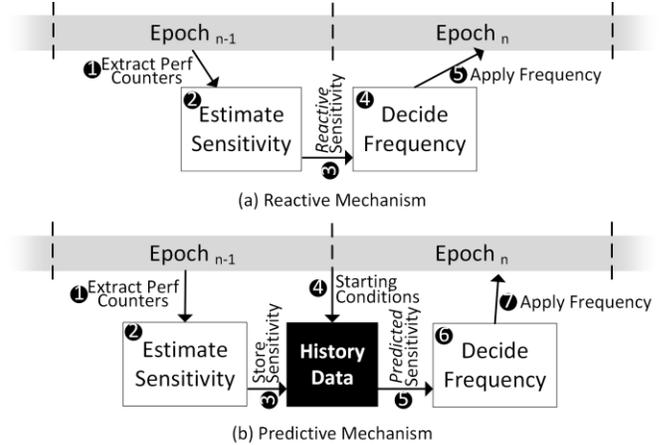

Figure 3: (a) Reactive DVFS mechanisms adopted in traditional approaches which assume that phases remain similar across consecutive time epochs. (c) Predictive mechanism approach which caters to the highly varying sensitivity.

**CRISP.** The CRISP model [20] extended the Critical Path model by considering the high number of store stalls observed in GPU, as well as the high memory-computation overlap. They calculate both core and memory time by selectively measuring store stalls and computation overlap. It is important to note that, CRISP assumes a CU to be equivalent to a CPU core and approximates execution within a CU to a single-threaded workload as shown in Figure 2(a). Further, CRISP assumes a fixed-time epoch but extrapolates power and execution time estimates.

To our knowledge, none of these prior CPU approach accounts for multiple threads executing on the same core, which is common in GPUs. The CRISP model extends the CPU model to GPU CU leading to low accuracies in estimation at fine-grain time scales as we will discuss later.

## 2.4 Predicting Sensitivity

The second part of the challenge, which has often been overlooked by previous works, is predicting the sensitivity of future time epochs. The performance modeling techniques discussed above mostly restrict themselves to reactive approaches. In other words, they estimate the performance of a work segment after executing it and then immediately apply the estimation to the next work segment or time epoch, as shown in Figure 3(a). Such reactive approaches are generally referred to as last-value predictors [55]. Longer-term value predictions, which predict the duration of a continuous phase, were also proposed [56]. Such reactive systems rely on the assumption that workloads exhibit similar behavior over consecutive work segments or time epochs [20, 10]. A few efforts that improve upon this assumption use a global phase history table to predict the variation across consecutive time epochs [55, 57]. The intuition behind such a table is that CPU workloads often exhibit small repetitive patterns in performance behavior. A recent work [4] utilizes a Q-learning mechanism to predict V/f state directly using a set of attributes.

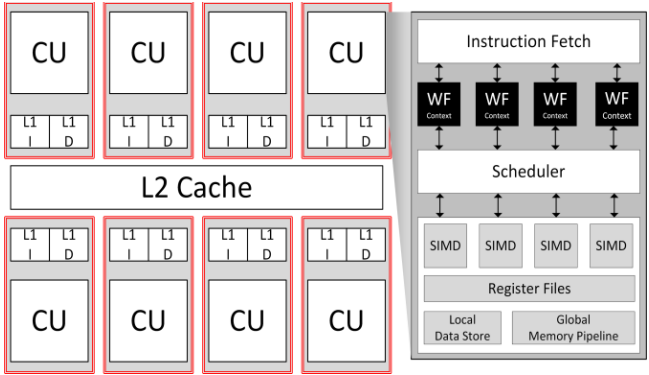

**Figure 4: A typical GPU system architecture with separate V/f domains (red lines) for each compute unit and L1 caches. The inset shows the microarchitecture of a CU.**

As we will see later in Section 3.3, these reactive policies are not sufficient for obtaining maximum power efficiency gains in the case of fine-grain DVFS in GPUs.

## 3 FINE-GRAIN VOLTAGE-FREQUENCY ISLANDS

To explore fine-grain DVFS, we consider an AMD Vega[TM] GPU [58] with multiple V/f domains as shown in Figure 4. Specifically, for most of our evaluations, we assume a fine-grain V/f domain that comprises of one CU along with its L1 caches. Later, Section 6.5 shows our single-CU observations apply to systems with multiple CUs in a V/f domain as well.

### 3.1 Fixed-Time Epoch

For DVFS domains tunable to different frequencies, even as low as 1 μs, it is imperative to manage the domains with a purely hardware-mechanism. A software-managed mechanism would be unable to react fast enough at such short timescales [31]. At these short time scales, we advocate for a fixed time-window-based approach for managing DVFS because the system can consistently adapt to the minor changes in the workload behavior with minimal overhead to IVRs and PLLs. In contrast, a fixed-instruction-window approach could either miss productive transition possibilities or encounter frequent unproductive transitions leading to resonance noise [59]. This effect is especially magnified for GPUs where there is high variation in instructions committed over time (as discussed later in this section). Thus, a fixed-time epoch control ensures that transitions occur quickly while amortizing the transition power overhead.

### 3.2 Characterization of Fine-Grain Phases

DVFS requires a quantitative characterization of the phase so that an optimal frequency can be chosen. Earlier works characterized workloads based on the sensitivity of execution time to operating frequency [55, 60]. Such characterization helps optimize DVFS at a thread-granularity to meet performance deadlines and minimize ED[n]P. Later works utilizing finer granularities of DVFS have focused on a fixed number of instructions [10, 20, 61, 62] and characterized execution time sensitivity in a similar manner. However, a different metric would be required to analyze the frequency sensitivity of fixed-time epochs. The general approach to measuring frequency sensitivity of different phases is to characterize them as compute-intensive—if the amount of work done increases substantially with frequency—or memory-bound—if otherwise.

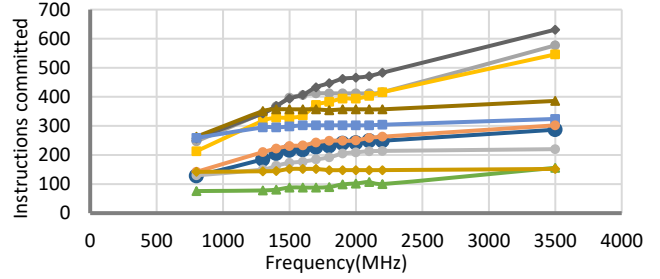

**Figure 5: Instructions committed in time epoch (1 μs) by a CU executing *CoMD*. Each set (color) of data points represents a unique sampled time epoch operated at various frequencies.**

To quantitatively measure the *work done* by a system, we consider the number of critical instructions committed for any given time epoch. We measure critical instructions as the instructions that lead to a memory operation or end of the kernel. This does not include the instructions which lead up to a memory or synchronization barrier. For a sampled set of unique time epochs (Section 5.1) in the application *comd*, Figure 5 shows the number of instructions committed at different operating frequencies. As expected for the compute-intensive epochs, the number of instructions committed increases with increases in frequency (high slope). Meanwhile for the memory-bound epochs, the increase is low or negligible (low slope).

The key observation in Figure 5 is that the number of instructions committed has a mostly linear relationship to operating frequencies for the range most appealing to fine-grain DVFS. Performing linear regression over the values obtained for different time epochs and workloads corroborated this observation with an average $R^2$ value of 0.82. This strong correlation lets us model the performance at each operating frequency using a linear model. In other words, we could model the number of instructions executed $I_f$ at frequency $f$ as:

$$I_f = I_0 + Sf$$

The term $I_0$ signifies the minimum number of instructions that would be executed. The term $S$ here quantifies the performance sensitivity of the fine-grain time epoch to the operating frequency. We define *Sensitivity* of a time epoch assuming a given starting condition. Sensitivity signifies the potential increase in instruction throughput for an increase in unit frequency. This term, therefore, quantifies the phase of the fine-grain workload; higher sensitivity values indicate a more compute-intensive work segment, while a lower value signifies a memory-intensive work segment.

$$Sensitivity = \frac{\Delta\ Instructions}{\Delta\ Frequency}$$

It is important to note that the linear model is suitable for our targeted range of GPU DVFS frequencies (1.0-3.0 GHz), but not all frequencies. We observed that frequencies as low as 0.8 GHz and as high as 3.5 GHz exhibit a considerably similar linear model as shown in Figure 5. However, at extreme low/high frequencies (e.g., 200 MHz), we expect this model to be insufficient.

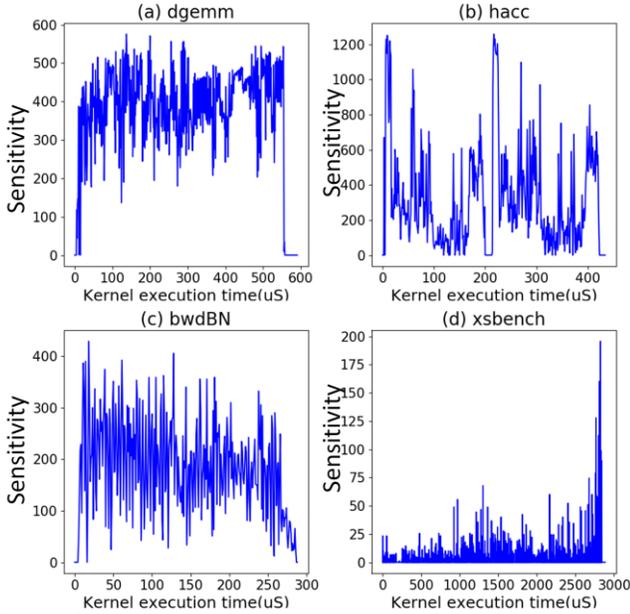

**Figure 6: Highly varying sensitivity profile of (a) *dgemm*, (b) *hacc*, (c) *BwdBN*, and (d) *xsbench* applications.**

## 3.3 Variability of Sensitivity in GPU Applications

The sensitivity metric defined above quantitatively characterizes the phase of any specific fine-grain time epoch. We profiled the sensitivities of GPU kernels over the course of their execution to understand the phased behavior at the fine-grain time scales. Each time epoch per V/f domain has its own sensitivity depending on the work-segment being executed.

We observed that the variation in sensitivity across consecutive time epochs in GPU workloads is extremely high. Figure 6(a)-(d) shows an example of how the sensitivity of the *several GPU* applications highly varies over time. Across a variety of workloads, Figure 7(a) quantifies the average relative sensitivity change in *consecutive time epochs*. We see that sensitivity varies by an average of 37% across consecutive 1μs time epochs. For highly compute-intensive workloads, this results in large differences in performance as frequency changes. In addition, the variation across time epochs increases as we go to finer-grain time epochs. Figure 7(b) shows that the average relative change across workloads increases from a 12% variation at 100μs time epochs to a 37% variation at 1μs. This observation counters the assumptions of most previous works that GPU behavior remains relatively consistent for consecutive time epochs. While such an assumption holds well for coarse-grain time epochs, our observations clearly show that it fails to adhere when it comes to fine-grain time scales.

The varied nature of GPU frequency sensitivity contradicts the conventional wisdom that reactive DVFS mechanisms can scale down to fine-grain time epochs. Instead, a predictive mechanism (Figure 3(b)) capable of accurately estimating the frequency sensitivity of future time epochs is required. Such a mechanism needs to consider the large variability of behavior across different kernels as well as behavior variance within each kernel from epoch to epoch. However, predicting such a highly varying sensitivity requires a deep understanding of the architecture involved. We take these challenges into account when designing our fine-grain DVFS prediction mechanism.

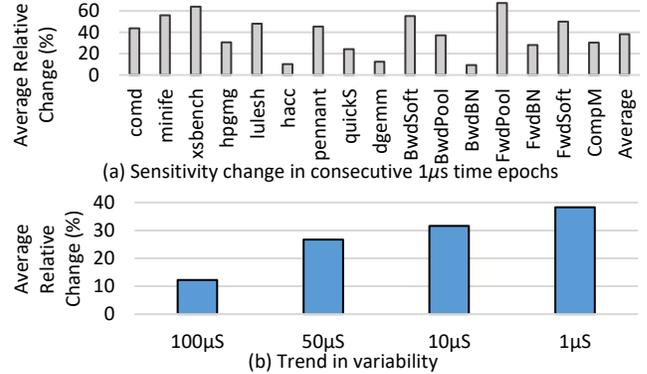

**Figure 7: (a) Average relative change in sensitivity of consecutive time epochs (1μs) in each CU for GPU workload. Several applications show a high variation across consecutive time epochs. (b) The average relative change increases as we go to finer-grain time epochs.**

## 4 FINE-GRAIN SENSITIVITY PREDICTION FOR GPUS

### 4.1 GPU Execution Hierarchy

Predicting performance in GPUs at different frequencies requires an understanding of their inherent execution hierarchy. Previous work [20] estimated GPU performance sensitivity by extending the CPU models to GPU CUs [10].

The underlying execution mechanism of a CU is, however, vastly different than the single-thread model assumed by prior GPU DVFS works. A CU executes many wavefronts (sometimes called warps) in a lock-step mechanism with individual program counters (PC). This results in an arbitrary mix of instructions being executed in any given time epoch. The mix of instructions is dependent on the individual progress of the wavefronts and could potentially be executed in any order (subject to individual wavefront ordering). This arbitrary mixture of instructions from different wavefronts thus determines the work segment of a CU in any given time epoch. Such an execution mechanism results in three levels of variation. First, the exact sequence of instructions committed by the CU in any given time epoch can highly vary for different operating frequencies. Second, because individual wavefronts progress at different rates, there is a high variation in instruction mixture and frequency sensitivity across consecutive time epochs. Last, each wavefront goes through different phases depending on the nature of their instructions and other environmental factors (memory traffic). These characteristics pose a challenge to the prediction of sensitivity for future time epochs. Figure 8 shows how wavefront-level variations within the sampled application *BwdBN* affect a CU's sensitivity across time epochs. Although some of these characteristics may exist in a multi-threaded CPU core, their effect is considerably lower because of the relatively low degree of multi-threading (2-8 threads) compared to a CU (approximately 40 waves).

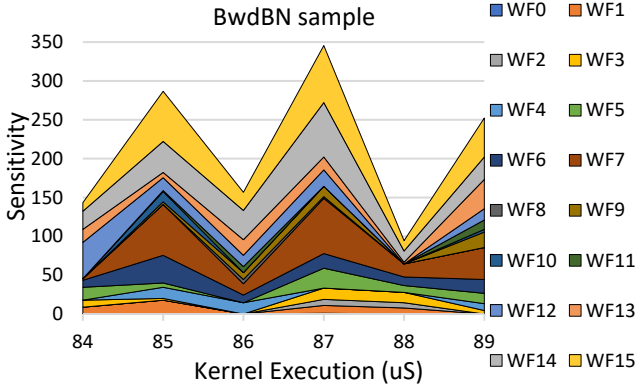

**Figure 8: Sample profile of a CU executing *BwdBN* application showing the contribution of wavefronts to the total sensitivity of the CU.**

### 4.2 Wavefront-Level Estimation

We propose utilizing wavefront-level sensitivity to estimate the total sensitivity experienced by a V/f domain. This requires estimating the sensitivity of resident wavefronts and aggregating them up as shown in Figure 2(b).

The execution of wavefronts is comparative to the execution of in-order CPU threads. Each wavefront has a PC denoting the next instruction to be executed and wavefronts execute and commit instructions in-order. Also, similar to multi-threaded CPU execution, when a wavefront stalls waiting for load dependencies, other ready wavefronts consume execution resources and progress. However, the wavefronts can interfere with each other because of unique GPU resource contention, such as explicit barriers and the CU's wavefront scheduling policy. However, due to their similarities to CPU threads, we apply the prior CPU DVFS models described in Section 2.3, to *estimate* per-wavefront sensitivity for any time epoch.

The next step is to aggregate the sensitivity obtained at the wavefront-level to the V/f-domain level. Because of the commutative nature of the sensitivity metric defined in Section 3.2, the sensitivity of a V/f domain would just be the sum of the sensitivities of the constituent CUs. The sensitivity of a CU itself would involve the combined sensitivities of the constituent wavefronts. Formally, this can be generalized to:

$$Sens_{Domain} = \frac{\Delta Ins}{\Delta Freq} = \frac{1}{\Delta Freq} \sum_{j=1}^{n\_CU} \sum_{i=1}^{n\_wf} \Delta Ins_{WF_{j,i}} = \sum_{j=1}^{n\_CU} \sum_{i=1}^{n\_wf} Sens_{WF_{j,i}}$$

Thus, having a scalable sensitivity metric helps us determine the performance of a V/f domain without losing detail from wavefront-level variations.

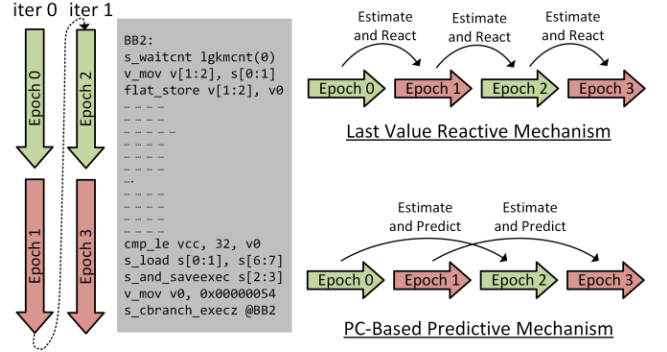

**Figure 9: PC-based phased prediction model leverages the repetitive nature of kernel execution. This example shows how later iterations leverage information from prior iterations. The estimate can either be used by later iterations within the same wavefront or different wavefront.**

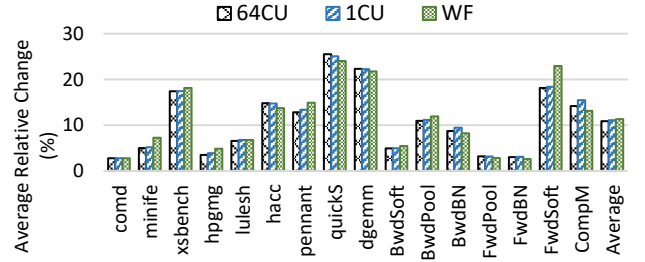

**Figure 10: Average relative change in CU sensitivity across consecutive iterations (1μs) starting from same respective wavefront PC address.**

### 4.3 Wavefront-Level Prediction

The main challenge for predicting GPU frequency sensitivity assuming fine-grain DVFS is their high variation, as discussed in Section 3.3. Due to the high variability, it is not sufficient to utilize a simple reactive mechanism. Instead, the mechanism must anticipate the variation in sensitivity beforehand.

**Wavefront Phase.** One of the key reasons for the high variation in frequency sensitivity is the independent progress of each wavefront. Moreover, the wavefronts themselves exhibit phases depending on the nature of instructions and environmental conditions. To observe the predictability of wavefronts, we measure the difference in wavefront-level sensitivity in consecutive iterations starting from the same starting PC address (Epoch 0 vs Epoch 2 in Figure 9). This study was designed to identify whether wavefronts would exhibit similar sensitivity when executing the same sequence of instructions. Figure 10 shows the relative change in sensitivity of wavefronts starting from any given PC over consecutive iterations. The different granularities (64CU, CU, WF) represent the cases when iterations within the respective boundaries were considered. We observe that the change in sensitivity over *consecutive iterations* of the same sequence of instructions is only 10% on average. This is much lower than the 37% average change observed for consecutive time epochs (Figure 7) and shows that the inherent sensitivity of a time epoch in any wavefront is primarily determined by the nature and order of the instructions executed.

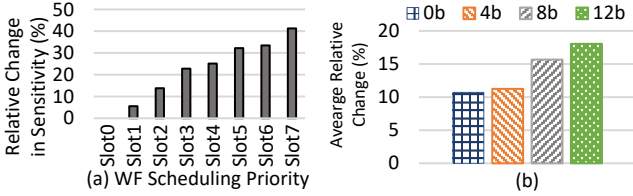

Figure 11: (a) Change in sensitivity (1 μs) observed for different wavefront slots in *quickS* (highest inter-wavefront variation). (b) Average relative change in sensitivity when different index offset values for PC-table with a CU-level granularity.

**Contention Across Wavefronts.** To understand the effect of contention across wavefronts, we compared the change in sensitivity for a wavefront to the highest priority wavefront. Figure 11(a) shows the average relative difference in sensitivity observed for different wavefront slots for *quickS* application. This shows that most of the difference in performance is because of contention between wavefronts within a CU; with the highest priority wavefront experiencing no impact on sensitivity and the lower priority wavefronts experiencing an increased relative change in sensitivity. We attribute this variation to the GPU's 'oldest-first' wavefront scheduling policy which prioritizes the execution of the oldest ready wavefront.

### 4.4 Novel PC-Based Phase Prediction Unit

The observations above motivated us to build PCSTALL, a wavefront-level PC-based sensitivity predictor as shown in Figure 9. The predictor leverages the repetitive behavior of instruction sequences across wavefronts or iterations to accurately predict sensitivity and is composed of two mechanisms: update and lookup. At the end of each epoch, each wavefront estimates their epoch's sensitivity and stores the estimate in a table (update mechanism). For the next epoch, the predictor accesses the same table using each wavefront's next PC to estimate the overall sensitivity (lookup mechanism). Such a PC-based predictor is more feasible in GPU kernels where the code size is often limited and iteratively executed by many wavefronts. These characteristics ensure that the table is quickly populated for successful retrievals.

**Estimation Model.** For wavefront-level estimation, we utilize the STALL model which measures the time spent by wavefronts stalled waiting for memory responses. The time spent stalled by a wavefront can be directly measured by the time spent blocked at the *s_waitcnt* instruction in AMD Vega™ ISA [58]. The remaining total core time spent by the wavefront is then used to estimate the sensitivity.

$$Sens_{WF} = IPC_{WF} \times T_{core,WF}$$

The estimated sensitivity is further normalized depending on the relative age (i.e., scheduling preference) of the wavefront. GPUs strive to schedule the oldest wavefront within the CU first, thus our STALL model considers the scheduling contention experienced by the wavefront in the overall sensitivity estimate. We explored further optimizations to the estimation model but, we found that this simple model is enough to achieve high accuracy when combined with the PC-based prediction model.

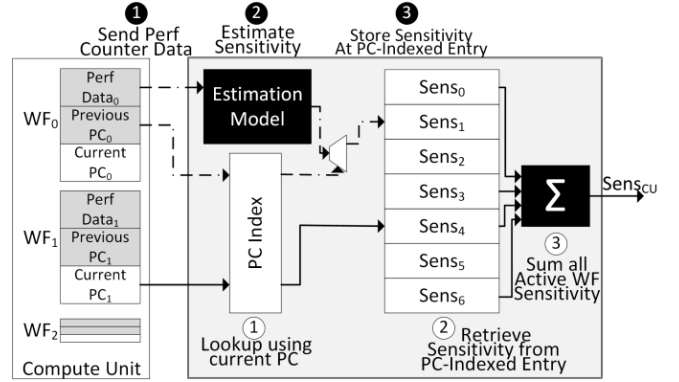

Figure 12: PCSTALL: Microarchitecture of PC-indexed sensitivity table. PC of wavefronts are used to index into a table that stores the information about the sensitivity of the time epoch starting from the PC. The update (*top dotted lines*) and lookup (bottom solid lines) mechanisms are also depicted.

**Lookup Mechanism.** ①Wavefronts index into a table using their current PC address to ② retrieve wavefront-level sensitivity of the upcoming time epoch. ③ The individual sensitivities are then summed up to calculate the overall sensitivity of a CU. The CU sensitivity is then used to predict instructions committed at different frequencies. We model the table such that wavefronts index into them one by one at a fixed cycle before the start of a time epoch. Thus, some latency is incurred in the lookup mechanism. This can be improved by further optimizations.

**Update Mechanism.** After the execution of the time epoch, ❶ each wavefront's sensitivity is calculated using the previously discussed estimation model. ❷ The estimated sensitivities are ❸ stored into the table for future reference. The update mechanism happens in a non-critical path and has no latency impact on future predictions.

**Hardware Design.** The PC-based predictor requires one or more tables to store the sensitivities and can be shared by multiple wavefronts. The negligible reduction in accuracy (inferred from Figure 10) when the table is shared across a different number of wavefronts provides flexibility on where the tables are placed.

Each wavefront needs to index into the table using the starting PC-address. For tuning the offset bits, we calculated the relative change in consecutive iterations for different offsets as shown in Figure 11(b). We observe that the relative change starts increasing when the PC-address offset is greater than 4 bits (~ 4 instructions per entry). For tuning the number of entries in PC-table we calculated the hit ratio at different sizes and observed that 128 entries were sufficient for achieving 95%+ hit ratio. Because most workloads involve loops of a few hundred instructions we set the PC table to 128 entries (covering 512 instructions). TABLE I presents the hardware storage overhead of our PC-based predictor per instance. PCSTALL consumes less storage and thus less power compared to the state-of-art CRISP model. The PC table could either be instantiated one per CU or shared among many CUs.

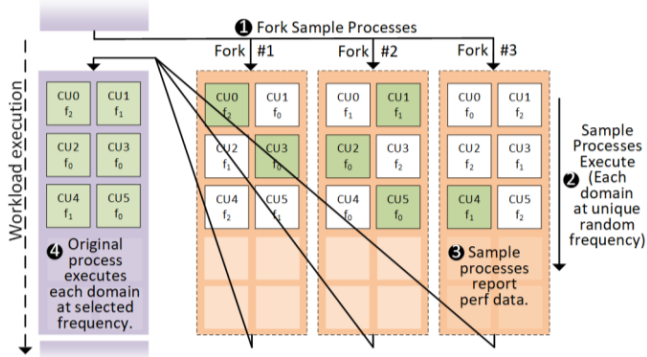

**Figure 13: Fork-Pre-Execute Methodology:** Simulator process is forked ❶ at each time epoch and sample processes are executed ❷ allotting a unique shuffled frequency (e.g., $f_0$, $f_1$, and $f_2$) for each V/f domain. Performance data is sent ❸ to the original process where optimal frequencies are then selected ❹ for each V/f domain and the time epoch is re-executed.

## 5 METHODOLOGY

We utilize the publicly available AMD GCN3 based GPU simulation model [58] within the gem5 simulator [63, 64] for evaluating the performance of the system. We assume a 64 CU GPU with 16 L2 banks shared among all CUs [65]. The V/f domain of the memory subsystem, along with L2 cache, is fixed at 1.6 GHz for all the evaluations. For most of our evaluations, we assume a single V/f domain for each CU, unless otherwise specified. Each V/f domain is DVFS-enabled [15] with an integrated voltage regulator and frequency modulator capable of transitioning to a frequency between 1.3 GHz – 2.2 GHz at steps of 100 MHz (10 V/f states). The transition latency is assumed to be 4ns for 1μs time epochs, 40ns for 10μs, 200ns for 200μs, and 400ns for 100μs time epochs. At the end of each epoch, the local DVFS manager uses the associated control mechanism to assign an operating frequency for each V/f domain for the next epoch subject to an objective function (EDP, $ED^nP$, etc.).

**Power Model.** For estimating the energy and power consumed by an application, we utilize an in-house power model based on detailed hardware measurements. The power model takes in performance counter data and estimates energy consumed for any given interval similar to previous works [66]. The model projects both dynamic and leakage power consumption across different voltage-frequency states. Note that, unlike dynamic power, the leakage power at the different operating states does not significantly vary across the small voltage range offered by the IVRs. The power model also accounts for the efficiency of IVRs at the different voltage states and the impact temperature has on leakage power. This baseline power model was validated against an AMD Radeon[TM] VII GPU. For our area estimates, we included the overheads for the IVRs [23], DLDOs [23], and V/f domain crossing units [67]. Our estimates show that the total area overhead for 1-CU V/f domain with PCSTALL each is about 4.59% of the overall GPU chip area.

**TABLE I. HARDWARE STORAGE OVERHEAD PER INSTANCE IN BYTES.**

| | | | | |
|---|---|---|---|---|
| **PCSTALL** | Sensitivity Table | 128 entries | 128 | |
| | Starting PC register (Only index bits) | 40x | 40 | **328** |
| | Stall Time Registers | 40x (1/WF) | 160 | |
| CRISP [20] | | | | 668 |
| CRIT [10] | | | | 660 |
| LEAD | | | | 18 |
| STALL | | | | 4 |

### 5.1 Oracle Simulation

Measuring a mechanism's accuracy is one of the key challenges when designing DVFS solutions. Earlier coarse-grain DVFS approaches [20] exhaustively sampled a workload across different frequencies to measure a method's accuracy. However, the fine-grain DVFS scenarios evaluated in this work presents two unique hurdles. First, with shorter time intervals, it is difficult to exhaustively simulate the same small segment at different frequencies, since each sample needs to start from the exact same starting conditions. Second, with multiple V/f domains operating simultaneously and contending for shared resources, the performance of a particular V/f domain not only depends on its operating frequency but on frequencies of other V/f domains as well. With deterministic simulation, the total possible execution paths for any given time epoch is given by $\#FreqStates^{\#ClockDomains}$. For our simulations, with 64 CUs and 10 possible V/f states, the number could be as high as $10^{64}$, making measurements intractable.

We solve these challenges by simplifying the search space and modifying the simulator to execute ahead, reporting back the performance of the pre-executed samples, before rolling back and executing the time segment a second time with the best-known frequencies for each domain. Figure 13 shows the process for determining an oracle measurement. We address the first search space challenge by forking the simulation process (parent) into multiple sampling processes (children). Each sampling process operates the V/f domains at a unique frequency. The sampling processes are executed for a time epoch before halting and sending their performance data to the original parent process where the time epoch is re-executed. The second V/f domain interference challenge is solved by shuffling the frequencies across the cores in the sampling processes. To validate our overall approach, we compare the per-domain performance reported by the sampling (pre-executed) processes to that of the original process re-executed at selected frequencies. As previously noted, a 100% accurate methodology would require $10^{64}$ processes. Instead, our solution reaches 97.6% accuracy with only 10 processes (one for each frequency state).

We use this mechanism to generate near-accurate estimations of any time epoch for a given V/f domain. In addition, we use accurate sensitivity estimates for a future time epoch for an oracular DVFS policy (ORACLE).

TABLE II. HPC AND MI WORKLOADS USED FOR EVALUATION AND THE NUMBER OF UNIQUE KERNELS IN BRACES.

| HPC Apps | MI Apps |
|---|---|
| Molecular Dynamics (*comd*) (1) | Double Prec. MatrixMul (*dgemm*) (1) |
| Full MultiGrid (*hpgmg*) (1) | Batch-Norm Back (*BwdBN*) (1) |
| Shock Hydrodynamics (*lulesh*) (27) | Pooling Backward (*BwdPool*) (1) |
| Finite Element (*minife*) (3) | Softmax Backward (*BwdSoft*) (1) |
| Monte Carlo Transport (*xsbench*) (1) | Batch-Norm Foreward (F*wdBN*) (1) |
| Cosmology Code (*hacc*) (2) | Pooling Forward (F*wdPool*) (1) |
| Monte Carlo Quicksilver (*quickS*) (1) | Softmax Forward (F*wdSoft*) (1) |
| Unstructured Mesh (*pennant*) (5) | |
| Discrete Ordinates (*snapc*) (1) | |

TABLE III. DVFS PREDICTION DESIGNS EVALUATED.

| Name | Estimation Model | Control Mechanism |
|---|---|---|
| STALL | Stall Model [24] | Reactive |
| LEAD | Leading Load [24, 32, 33] | Reactive |
| CRIT | Critical Path [10] | Reactive |
| CRISP | CRISP GPU Model [20] | Reactive |
| ACCREAC | Accurate Estimate | Reactive |
| **PCSTALL** | **Stall – Wavefront** | **PC-Based** |
| ACCPC | Accurate Estimate | PC-Based |
| ORACLE | Accurate Estimate | Oracle |

## 5.2 Objective Functions

The fundamental goal of DVFS is to improve energy efficiency. However, the exact objective function, that DVFS optimizes for, varies depending on the product, use-case, workload, and environment. Some system-level objective functions include minimizing $ED^nP$ metrics or maintaining operation within certain power bounds. Our evaluations consider minimizing EDP and $ED^2P$ because EDP is often important for battery-constrained environments while $ED^2P$ is important for performance-oriented servers. Our DVFS prediction mechanism could easily be extended to other objective functions such as meeting per-job quality-of-service (QoS) deadlines. Note that the prediction of performance will only tell us how the V/f domain will perform at different operating frequencies; choosing the appropriate frequency depends on the objective function and is orthogonal to the prediction mechanism.

Alternatively, one could also combine the two steps of prediction and frequency selection into a single step. However, our approach of implementing a prediction mechanism in an objective-agnostic manner allows the DVFS mechanism to adjust to various power-constrained, multi-V/f domain scenarios where one cannot directly assume any objective policy at design time.

## 5.3 Workloads and Baseline Models

We evaluate our approach using HPC and machine intelligence GPU workloads (TABLE II). Specifically, for HPC applications, we consider the ECP proxy applications [68] and for machine intelligence applications [69], we evaluate the DeepBench [70] and DNNMark [71] benchmark suites.

We compare our DVFS approach to the baseline models described in Section 2.3, three static frequencies, as well as the oracle discussed earlier in Section 5.1. TABLE III lists all evaluated designs including their estimation models and prediction mechanisms. We include three models that use the accurate estimates from our fork and pre-execute simulation methodology. The first uses the accurate estimates of the prior time epoch to create a reactive predictor (ACCREAC) and the second uses the accurate estimates to fill in a table for the accurate, but not practical, PC-based predictor (ACCPC). Finally, we directly use the accurate estimates for the next time epochs in the ORACLE model for near-optimal comparison.

Our PC-based predictor could potentially be combined with any of the previously described wavefront-level estimation models, but we chose the STALL model because of its relatively simplicity. As described above, we modified the STALL model to directly estimate the sensitivity of each wavefront, in contrast to directly applying it at a CU-level [20] or GPU-level.

## 5.4 Hierarchical Power Management

The hardware based DVFS system described in this paper has been designed with a commercial hierarchical power management system in mind. Within such a scheme, higher-level power management policies set power objectives at millisecond scales, which then impact the internal frequency range used by the hardware DVFS controller. For our evaluations, we chose a small range of frequencies (1.3GHz-2.2GHz) to simulate the power constraint set by a higher-level power manager above.

## 6 EVALUATIONS

Measuring the advantages achieved by our DVFS prediction approach involves assessing the accuracy of the predictor as well as the overall power efficiency improvement. In this section, we first evaluate the accuracy of the predictor by comparing it to the oracle. Then we present results optimizing for minimal $ED^2P$ and EDP.

### 6.1 Prediction Accuracy

We calculate prediction accuracy by comparing the number of predicted instructions committed to the number of actual instructions committed. Note that the prediction accuracy is a power-model-agnostic metric and only focuses on the prediction algorithm itself. Figure 14 shows the prediction accuracy of different models compared to an oracle reported sensitivity at 1μs. We observe that complex models such as CRIT and CRISP outperform simpler models such as STALL and LEAD yet still have a relatively low prediction accuracy of ~60%. Most of this inaccuracy is due to the reactive nature of prior approaches. Even with an accurate sensitivity estimation, a reactive model (ACCREAC) is only able to reach 63% accuracy in prediction on an average.

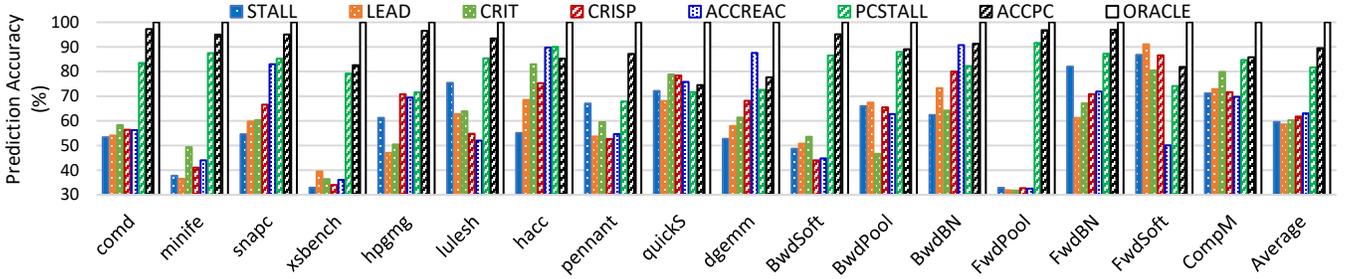

Figure 14: Accuracy of different prediction models compared to an ORACLE prediction (100% accurate) at 1μs time epochs. A practical implementation of a PC-based predictor (PCSTALL) outperforms even a perfect reactive model (ACCREAC) in almost all the models.

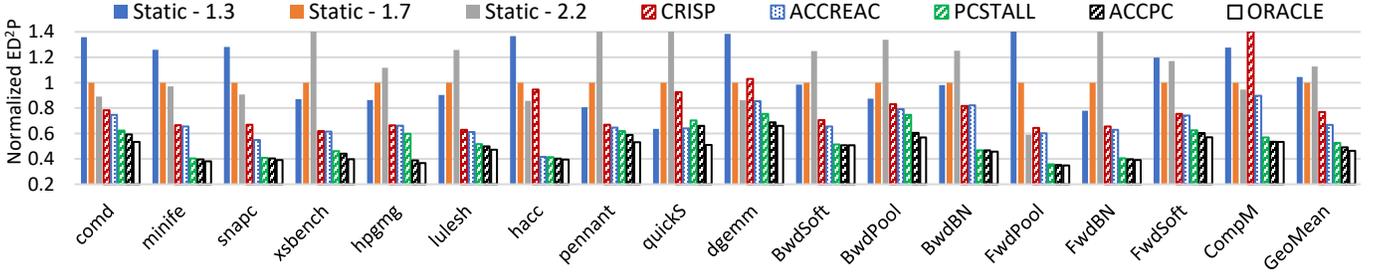

Figure 15: ED$^2$P of workloads normalized to a static 1.7GHz execution at 1μs time epochs.

On the other hand, a PC predictor has the potential (ACCPC) of reaching 90% prediction accuracy. Even with a practical sensitivity estimation model, our PC-predictor (PCSTALL) can deliver up to 81% prediction accuracy, outperforming the accurately estimating reactive model (ACCREAC). These results show that high prediction accuracies can be achieved even with a simple prediction mechanism. These prediction accuracies directly translate into better frequency selection and result in improved power efficiencies as shown in the next subsections.

Figure 1(b) shows the trend of accuracy with different time epoch durations. Although PCSTALL delivers the best accuracies at finer time epochs, PCSTALL also results in improvement at longer time epochs. Our results show that PCSTALL can be beneficial even at 50μs time epochs, showing that our policy can be adopted in current generation hardware as well.

### 6.2 Minimizing ED$^2$P

Next, we evaluate how the different models minimize overall ED$^2$P. Figure 15 shows the ED$^2$P values normalized to a static 1.7 GHz operation. The ORACLE improves power efficiency by up to 54%, whereas the reactive models provide just a 34% potential improvement with CRISP delivering 23% improvement in ED$^2$P compared to 1.7 GHz operation. On the other hand, our PC-based predictive model showcases potential savings in ED$^2$P of 48% while ACCPC can reach up to 51% improvement.

Figure 1(a) shows the trend of improvement in ED$^2$P when the DVFS time epoch is reduced from 100μs to 1μs. We see that the advantage of PCSTALL improves as we approach finer time epochs. Even at a 50μs time epoch, PCSTALL delivers a 36% improvement compared to a 21% improvement by CRISP.

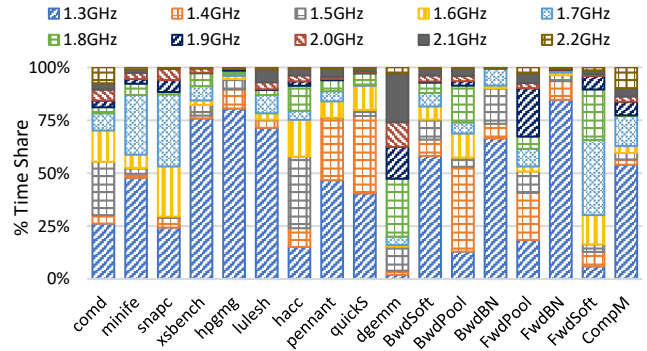

Figure 16: Average time share of each frequency state in CUs executing the different workloads while optimizing for ED$^2$P using PCSTALL at 1μs.

These results reinforce the intuition that a future-looking mechanism, rather than a reactive one, is required to achieve high prediction accuracy of the next epoch's sensitivity and consequently, achieve better power-performance at fine-grain DVFS time epochs.

**Frequency Time Share.** Figure 16 shows the percentage of time CUs spend at each frequency state using the PCSTALL mechanism for ED$^2$P optimization. As expected, the CUs frequently select the higher operating frequencies for the compute-intensive applications (like *dgemm* and *hacc*), while the CUs mostly stay within the lower frequency ranges for the memory-intensive applications (like *hpgmg* and *xsbench*). Workload *dgemm* also has a highly heterogeneous behavior, leading to comparatively lower accuracies.

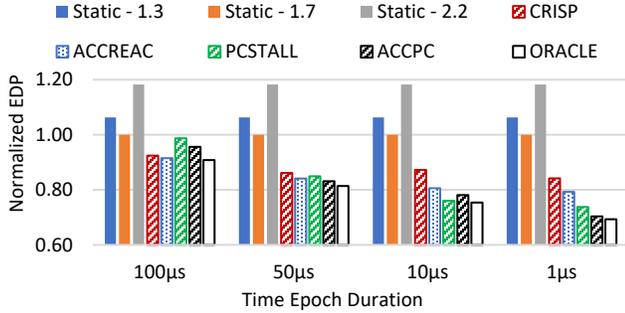
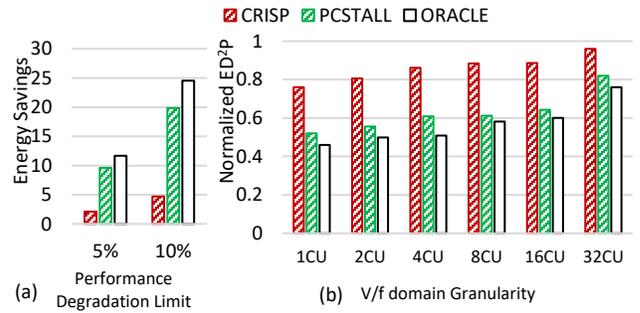

Figure 17: Geomean EDP of workloads normalized to a static 1.7 GHz execution at different time epoch durations.

Figure 18: (a) Average energy savings with prediction mechanisms at different performance degradation limits. (b) Average normalized $ED^2P$ for different V/f domain granularity in a 64CU GPU.

*BwdPool* shows an interesting behavior where it adopts a single frequency (1.5 GHz) during its training period. This is mainly because of the constant rate of instruction executed by the CUs during its execution.

*FwdSoft* behaves in a special way where a static operation at 1.7 GHz is beneficial over both 1.3GHz and 2.2GHz. Our detailed analysis showed that this is because of a second-order effect where higher frequency operation of multiple CUs leads to cache thrashing at the second level of cache. This leads to performance degradation at 2.2 GHz.

### 6.3 Minimizing EDP

Figure 17 shows the trend in EDP improvement delivered by PCSTALL. The results show a similar trend as $ED^2P$. While PCSTALL is able to take advantage of the finer-grain time epochs, reactive policies fail to show a considerable advantage. However, the difference in reactive and predictive policies is lower compared to $ED^2P$ optimizing executions.

### 6.4 Fixed-Performance Energy Savings

In addition to the $ED^2P$ and EDP objective functions, we studied the overall energy savings within different performance degradation limits. Figure 18(a) shows the energy savings with 5% and 10% performance degradation limits. We see that PCSTALL can deliver up to 9.6% energy savings when restricted to 5% performance degradation compared to 2.1% savings from CRISP. The difference between PCSTALL and CRISP increases when the degradation limit is increased to 10% where PCSTALL results in 19.86% energy savings compared to a 4.7% energy saving from CRISP.

### 6.5 Scalability Study

In addition to the single CU V/f domains, we study the scalability of PCSTALL for larger V/f domains. Figure 18(b) shows the normalized $ED^2P$ improvement delivered by PCSTALL at different V/f domain granularity in comparison to CRISP and ORACLE. Note that increasing the V/f domain granularity would mean fewer IVRs and shared PC-tables, with associated power and energy costs.

In general, as the V/f domain granularity increases, the opportunity in reducing the $ED^2P$ via DVFS decreases because of the decrease in opportunities to customize V/f state for the instructions being executed. In comparison to ORACLE, CRISP does not extract much $ED^2P$ improvements, resulting in only a 4% improvement over a static operation at 32CU granularity. On the other hand, PCSTALL delivers high improvement in $ED^2P$ even at the 32CU V/f domain granularity reaching 18% improvement compared to 24% improvement possible with ORACLE.

This shows that PCSTALL could be beneficial not just for smaller V/f domains but for larger domains as well. The scalable nature of PCSTALL enables high power efficiency gains leveraging the continuous improvement in circuit technologies.

## 7 CONCLUSION

In this paper, we presented prediction-based techniques to unlock the fine-grain DVFS opportunities provided by the improving IVR technology. We targeted GPUs as they operate over wider dynamic voltage ranges and thus have a higher potential for power savings. Our key observation was, at fine-grain time scales, GPU workloads exhibit high variation in sensitivity with respect to the operating clock frequency. Further, we observed that existing performance estimation models work more accurately at a wavefront level than at a CU or GPU level. Combining these observations, we advocated for a predictive model more suited to rapidly adjust to highly varying workload behavior and designed a scalable wavefront-level PC-based predictor that accurately predicts the frequency sensitivity of the future time interval. Our analysis showed that our prediction-based mechanism significantly outperforms previously proposed reactive techniques and comes close to achieving oracular prediction accuracies resulting in high energy efficiencies.

### ACKNOWLEGEMENTS

The authors would like to thank several reviewers for their constructive feedback and suggestions for this paper. We also thank Gabriel Loh and Karen Prairie for their feedback on improving early drafts of this work.

# REFERENCES

[1] S. Kaxiras and M. Martonosi, "Computer Architecture Techniques for Power-Efficiency," Synthesis Lectures on Computer Architecture, vol. 3, no. 1, pp. 1-207, 2008.

[2] S. Eyerman and L. Eeckhout, "Fine-grained DVFS using on-chip regulators," ACM Transactions on Architecture and Code Optimization, vol. 8, no. 1, pp. 1-24, 2011.

[3] W. Kim, M. S. Gupta, G.-Y. Wei and D. M. Brooks, "System level analysis of fast, per-core DVFS using on-chip switching regulators," 2008. [Online]. Available: http://eecs.harvard.edu/~dbrooks/kim2008_hpca.pdf. [Accessed 25 6 2019].

[4] Y. Bai, V. W. Lee and E. Ipek, "Voltage Regulator Efficiency Aware Power Management," Sigplan Notices, vol. 45, no. 1, pp. 825-838, 2017.

[5] A. Paul, S. P. Park, S. Dinesh, Y. M. Kim, N. Borkar, U. R. Karpuzcu and C. H. Kim, "System-Level Power Analysis of a Multicore Multipower Domain Processor With ON-Chip Voltage Regulators," IEEE Transactions on Very Large Scale Integration (VLSI) Systems, vol. 24, pp. 3468-3476, 2016.

[6] E. A. Burton, G. Schrom, F. Paillet, J. P. Douglas, W. J. Lambert, K. Radhakrishnan and M. J. Hill, "FIVR — Fully integrated voltage regulators on 4th generation Intel® Core™ SoCs," 2014. [Online]. Available: https://ieeexplore.ieee.org/document/6803344. [Accessed 21 11 2019].

[7] M. Fojtik, B. Keller, A. Klinefelter, N. Pinckney, S. G. Tell, B. Zimmer, T. Raja, K. Zhou, W. J. Dally and B. Khailany, "A Fine-Grained GALS SoC with Pausible Adaptive Clocking in 16 nm FinFET," in 2019 25th IEEE International Symposium on Asynchronous Circuits and Systems (ASYNC), 2019.

[8] X. Sun, A. Boora, W. Zhang, V. R. Pamula and V. Sathe, "14.5 A 0.6-to-1.1V Computationally Regulated Digital LDO with 2.79-Cycle Mean Settling Time and Autonomous Runtime Gain Tracking in 65nm CMOS," IEEE International Solid- State Circuits Conference - (ISSCC), 2019.

[9] Y. Okuma, K. Ishida, Y. Ryu, X. Zhang, P.-H. Chen, K. Watanabe, M. Takamiya and T. Sakurai, "0.5-V input digital LDO with 98.7% current efficiency and 2.7-µA quiescent current in 65nm CMOS," 2010. [Online]. Available: http://lowpower.iis.u-tokyo.ac.jp/paper/2010_22.pdf. [Accessed 25 11 2019].

[10] R. Miftakhutdinov, E. Ebrahimi and Y. N. Patt, "Predicting Performance Impact of DVFS for Realistic Memory Systems," IEEE Micro, pp. 155-165, 2012.

[11] P. Paternoster, A. Maki, A. Hernandez, M. Grossman, M. Lau, D. Sutherland and A. Mathad, "XBOX Series X: A Next-Generation Gaming Console SoC," in 2021 IEEE International Solid- State Circuits Conference (ISSCC), 2021.

[12] P. A. Meinerzhagen, C. Tokunaga, A. Malavasi, V. Vaidya, A. Mendon, D. A. Mathaikutty, J. P. Kulkarni, C. Augustine, M. Cho, S. T. Kim, G. E. Matthew, R. Jain, J. F. Ryan, C.-C. Peng, S. Paul, S. R. Vangal, B. P. Esparza, L. Cuellar, M. Woodman, B. Iyer, S. Maiyuran, G. N. Chinya, C. Zou, Y. Liao, K. Ravichandran, H. Wang, M. M. Khellah, J. W. Tschanz and V. De, "An energy-efficient graphics processor featuring fine-grain DVFS with integrated voltage regulators, execution-unit turbo, and retentive sleep in 14nm tri-gate CMOS," 2018. [Online]. Available: https://ieeexplore.ieee.org/stamp/stamp.jsp?tp=&arnumber=8310172. [Accessed 16 8 2019].

[13] T. Singh, A. Schaefer, S. Rangarajan, D. John, C. Henrion, R. Schreiber, M. Rodriguez, S. Kosonocky, S. D. Naffziger and A. Novak, "Zen: An Energy-Efficient High-Performance X86 Core," IEEE Journal of Solid-state Circuits, vol. 53, no. 1, pp. 102-114, 2018.

[14] T. Burd, N. Beck, S. White, M. Paraschou, N. Kalyanasundharam, G. Donley, A. D. Smith, L. D. Hewitt and S. D. Naffziger, ""Zeppelin": An SoC for Multichip Architectures," IEEE Journal of Solid-state Circuits, vol. 54, no. 1, pp. 133-143, 2019.

[15] D. Bouvier, J. Gibney, A. Branover and S. Arora, "AMD Raven-Ridge APU: Delivering a new level of visual performance in an SoC," [Online]. Available: https://www.hotchips.org/hc30/1conf/1.05_AMD_APU_AMD_Raven_HotChips30_Final.pdf.

[16] B. Zimmer, Y. Lee, A. Puggelli, J. Kwak, R. Jevtic, B. Keller, S. Bailey, M. Blagojevic, P.-F. Chiu, H.-P. Le, P.-H. Chen, N. Sutardja, R. Avizienis, A. Waterman, B. Richards, P. Flatresse, E. Alon, K. Asanovic and B. Nikolic, "A RISC-V vector processor with tightly-integrated switched-capacitor DC-DC converters in 28nm FDSOI," 2015. [Online]. Available: https://ieeexplore.ieee.org/stamp/stamp.jsp?tp=&arnumber=7231305. [Accessed 12 2 2020].

[17] R. Muthukaruppan, T. Mahajan, H. Krishnamurthy, S. Mangal, A. Dhanshekhar, R. Ghayal and V. De, "A digitally controlled linear regulator for per-core wide-range DVFS of atom™ cores in 14nm tri-gate CMOS featuring non-linear control, adaptive gain and code roaming," in ESSCIRC 2017 - 43rd IEEE European Solid State Circuits Conference, Leuven, 2017.

[18] S. J. Kim, D. Kim, H. Ham, J. Kim and M. Seok, "A 67.1-ps FOM, 0.5-V-Hybrid Digital LDO With Asynchronous Feedforward Control Via Slope Detection and Synchronous PI With State-Based Hysteresis Clock Switching," 2018. [Online]. Available: https://ieeexplore.ieee.org/stamp/stamp.jsp?tp=&arnumber=8490730. [Accessed 12 2 2020].

[19] H. Li, X. Wang, J. Xu, Z. Wang, R. K. V. Maeda, Z. Wang, P. Yang, L. H. K. Duong and Z. W. Wang, "Energy-Efficient Power Delivery System Paradigms for Many-Core Processors," IEEE Transactions on Computer-Aided Design of Integrated Circuits and Systems, vol. 36, no. 3, pp. 449-462, 2017.

[20] R. Nath and D. M. Tullsen, "The CRISP performance model for dynamic voltage and frequency scaling in a GPGPU," in 2015 48th Annual IEEE/ACM International Symposium on Microarchitecture (MICRO), 2015.

[21] X. Mei, L. S. Yung, K. Zhao and X. Chu, "A measurement study of GPU DVFS on energy conservation," 2013. [Online]. Available: http://comp.hkbu.edu.hk/~chxw/papers/hotpower_2013.pdf. [Accessed 15 8 2019].

[22] J. Leng, A. Buyuktosunoglu, R. Bertran, P. Bose, Y. Zu and V. J. Reddi, "Predictive Guardbanding: Program-Driven Timing Margin Reduction for GPUs," IEEE Transactions on Computer-Aided Design of Integrated Circuits and Systems, vol. 40, no. 1, pp. 171-184, 2021.

[23] S. T. Kim, Y.-C. Shih, K. Mazumdar, R. Jain, J. F. Ryan, C. Tokunaga, C. Augustine, J. P. Kulkarni, K. Ravichandran, J. W. Tschanz, M. M. Khellah and V. De, "Enabling Wide Autonomous DVFS in a 22 nm Graphics Execution Core Using a Digitally

[23] Controlled Fully Integrated Voltage Regulator," IEEE Journal of Solid-state Circuits, vol. 51, no. 1, pp. 18-30, 2016.

[24] G. Keramidas, V. Spiliopoulos and S. Kaxiras, "Interval-based models for run-time DVFS orchestration in superscalar processors," 2010. [Online]. Available: http://pages.cs.wisc.edu/~kaxiras/papers/cf10_dvfs_model.pdf. [Accessed 25 6 2019].

[25] C. Isci, G. Contreras and M. Martonos, Live, Runtime Phase Monitoring and Prediction on Real Systems with Application to Dynamic Power Management, 2006, p. 359–370.

[26] K. Straube, J. Lowe-Power, C. Nitta, M. Farrens and V. Akella, "Improving Provisioned Power Efficiency in HPC Systems with GPU-CAPP," 2018 IEEE 25th International Conference on High Performance Computing (HiPC), 2018.

[27] C. Nugteren, G.-J. v. d. Braak and H. Corporaal, "Roofline-aware DVFS for GPUs," 2014. [Online]. Available: http://ece.neu.edu/groups/nucar/nucartalks/roofline-aware_dvfs_for_gpus.pdf. [Accessed 15 8 2019].

[28] A. Mishra and N. Khare, "Analysis of DVFS Techniques for Improving the GPU Energy Efficiency," 2015. [Online]. Available: http://file.scirp.org/pdf/ojee_2015121415504865.pdf. [Accessed 15 8 2019].

[29] Z. Tang, Y. Wang, Q. Wang and X. Chu, "The Impact of GPU DVFS on the Energy and Performance of Deep Learning: An Empirical Study," in Proceedings of the Tenth ACM International Conference on Future Energy Systems, Phoenix, 2019.

[30] B. Dutta, V. Adhinarayanan and W.-c. Feng, "GPU power prediction via ensemble machine learning for DVFS space exploration," 2018. [Online]. Available: https://vtechworks.lib.vt.edu/handle/10919/81997. [Accessed 15 8 2019].

[31] K. Fan, B. Cosenza and B. Juurlink, "Predictable GPUs Frequency Scaling for Energy and Performance," in Proceedings of the 48th International Conference on Parallel Processing, Kyoto, Japan, 2019.

[32] S. Eyerman and L. Eeckhout, "A Counter Architecture for Online DVFS Profitability Estimation," IEEE Transactions on Computers, vol. 59, no. 11, pp. 1576-1583, 2010.

[33] B. Rountree, D. K. Lowenthal, M. Schulz and B. R. d. Supinski, "Practical performance prediction under Dynamic Voltage Frequency Scaling," 2011. [Online]. Available: http://yadda.icm.edu.pl/yadda/element/bwmeta1.element.ieee-000006008553. [Accessed 25 6 2019].

[34] C. Isci, A. Buyuktosunoglu, C.-Y. Chen, P. Bose and M. Martonosi, "An Analysis of Efficient Multi-Core Global Power Management Policies: Maximizing Performance for a Given Power Budget," IEEE Micro, pp. 347-358, 2006.

[35] H. Hoffmann and M. Maggio, "PCP: A Generalized Approach to Optimizing Performance Under Power Constraints through Resource Management," 2014. [Online]. Available: https://usenix.org/system/files/conference/icac14/icac14-paper-hoffman.pdf. [Accessed 23 7 2019].

[36] S. Park, J. Park, D. Shin, Y. Wang, Q. Xie, M. Pedram and N. Chang, "Accurate Modeling of the Delay and Energy Overhead of Dynamic Voltage and Frequency Scaling in Modern Microprocessors," IEEE Transactions on Computer-Aided Design of Integrated Circuits and Systems, vol. 32, no. 5, pp. 695-708, 2013.

[37] V. Spiliopoulos, S. Kaxiras and G. Keramidas, "Green governors: A framework for Continuously Adaptive DVFS," 2011. [Online]. Available: http://diva-portal.org/smash/record.jsf?pid=diva2:474791. [Accessed 25 6 2019].

[38] M. Weiser, B. B. Welch, A. J. Demers and S. Shenker, "Scheduling for reduced CPU energy," 1994. [Online]. Available: https://link.springer.com/chapter/10.1007/978-0-585-29603-6_17. [Accessed 25 6 2019].

[39] P. Zou, A. Li, K. Barker and R. Ge, "Indicator-Directed Dynamic Power Management for Iterative Workloads on GPU-Accelerated Systems," in 2020 20th IEEE/ACM International Symposium on Cluster, Cloud and Internet Computing (CCGRID), 2020.

[40] A. Zou, J. Leng, Y. Zu, T. Tong, V. J. Reddi, D. M. Brooks, G.-Y. Wei and X. Zhang, "Ivory: Early-Stage Design Space Exploration Tool for Integrated Voltage Regulators," 2017. [Online]. Available: https://dl.acm.org/citation.cfm?id=3062268. [Accessed 23 7 2019].

[41] A. Paul, S. P. Park, D. Somasekhar, Y. M. Kim, N. Borkar, U. R. Karpuzcu and C.H.Kim, "System-Level Power Analysis of a Multicore Multipower Domain Processor With ON-Chip Voltage Regulators," IEEE Transactions on Very Large Scale Integration (VLSI) Systems, 2016.

[42] M. Sankarasubramanian, K. Radhakrishnan, Y. Min, W. Lambert, M. J. Hill, A. Dani, R. Mesch, L. Wojewoda, J. Chavarria and A. Augustine, "Magnetic Inductor Arrays for Intel® Fully Integrated Voltage Regulator (FIVR) on 10th generation Intel® Core™ SoCs," in 2020 IEEE 70th Electronic Components and Technology Conference (ECTC), Orlando, 2020.

[43] C. Schaef, K. Radhakrishnan, K. Ravichandran, J. W. Tschanz, V. De, N. Desai, H. K. Krishnamurthy, X. Liu, K. Z. Ahmed, S. Kim, S. Weng, H. T. Do and W. J. Lambert, "A Light-Load Efficient Fully Integrated Voltage Regulator in 14-nm CMOS With 2.5-nH Package-Embedded Air-Core Inductors," IEEE Journal of Solid-state Circuits, vol. 54, no. 12, pp. 3316-3325, 2019.

[44] G. Yan, Y. Li, Y. Han, X. Li, M. Guo and X. Liang, "AgileRegulator: A hybrid voltage regulator scheme redeeming dark silicon for power efficiency in a multicore architecture," 2012. [Online]. Available: http://cs.sjtu.edu.cn/~guo-my/pdf/conferences/c130.pdf. [Accessed 5 7 2019].

[45] Q. Deng, D. Meisner, A. Bhattacharjee, T. F. Wenisch and R. Bianchini, "CoScale: Coordinating CPU and Memory System DVFS in Server Systems," IEEE Micro, pp. 143-154, 2012.

[46] S. Akram, J. B. Sartor and L. Eeckhout, "DVFS performance prediction for managed multithreaded applications," 2016. [Online]. Available: https://biblio.ugent.be/publication/7245653. [Accessed 23 7 2019].

[47] H. Zhang and H. Hoffmann, "Maximizing Performance Under a Power Cap: A Comparison of Hardware, Software, and Hybrid Techniques," Sigplan Notices, vol. 44, no. 2, pp. 545-559, 2016.

[48] H. Hoffmann and M. Maggio, "PCP: A Generalized Approach to Optimizing Performance Under Power Constraints through Resource Management," 2014. [Online]. Available: https://lup.lub.lu.se/search/publication/44544051-23aa-4be2-904f-b780181c3f90. [Accessed 5 7 2019].

[49] C. C. Lin, C. J. Chang, Y. C. Syu, J. J. Wu, P. Liu, P. W. Cheng and W. T. Hsu, "An Energy-Efficient Task Scheduler for Multi-

core Platforms with Per-core DVFS Based on Task Characteristics," 2014. [Online]. Available: http://ieeexplore.ieee.org/document/6957247. [Accessed 5 7 2019].

[50] M. Curtis-Maury, J. Dzierwa, C. D. Antonopoulos and D. S. Nikolopoulos, "Online power-performance adaptation of multithreaded programs using hardware event-based prediction," 2006. [Online]. Available: http://people.cs.vt.edu/~dsn/papers/ics06.pdf. [Accessed 5 7 2019].

[51] B. Su, J. Gu, L. Shen, W. Huang, J. L. Greathouse and Z. Wang, "PPEP: Online Performance, Power, and Energy Prediction Framework and DVFS Space Exploration," IEEE Micro, pp. 445-457, 2014.

[52] S. Bharadwaj, S. Das, Y. Eckert, M. Oskin and T. Krishna, "DUB: Dynamic Underclocking and Bypassing in NoCs for Heterogeneous GPU Workloads," in 2021 15th IEEE/ACM International Symposium on Networks-on-Chip (NOCS), 2021.

[53] Q. Wu, M. Martonosi, D. W. Clark, V. J. Reddi, D. A. Connors, Y. Wu, J. Lee and D. M. Brooks, "A Dynamic Compilation Framework for Controlling Microprocessor Energy and Performance," IEEE Micro, pp. 271-282, 2005.

[54] S. Akram, J. B. Sartor and L. Eeckhout, "DEP+BURST: Online DVFS Performance Prediction for Energy-Efficient Managed Language Execution," IEEE Transactions on Computers, vol. 66, no. 4, , pp. pp. 601-615, 1 April 2017.

[55] C. Isci, G. Contreras and M. Martonosi, "Live, Runtime Phase Monitoring and Prediction on Real Systems with Application to Dynamic Power Management," IEEE Micro, pp. 359-370, 2006.

[56] C. Isci, A. Buyuktosunoglu and M. Martonosi, "Long-term workload phases: duration predictions and applications to DVFS," IEEE Micro, vol. 25, no. 5, pp. 39-51, 2005.

[57] W. L. Bircher and L. K. John, "Predictive power management for multi-core processors," , 2010. [Online]. Available: https://link.springer.com/chapter/10.1007/978-3-642-24322-6_21. [Accessed 21 11 2019].

[58] A. Gutierrez, B. M. Beckmann, A. Dutu, J. Gross, M. LeBeane, J. Kalamatianos, O. Kayiran, M. Poremba, B. Potter, S. Puthoor, M. D. Sinclair, M. Wyse, J. Yin, X. Zhang, A. Jain and T. G. Rogers, "Lost in Abstraction: Pitfalls of Analyzing GPUs at the Intermediate Language Level," 2018. [Online]. Available: https://ieeexplore.ieee.org/document/8327041. [Accessed 30 7 2019].

[59] R. Thomas, N. Sedaghati and R. Teodorescu, "EmerGPU: Understanding and mitigating resonance-induced voltage noise in GPU architectures," 2016. [Online]. Available: https://ieeexplore.ieee.org/document/7482076. [Accessed 24 11 2021].

[60] C. Isci and M. Martonosi, "Phase characterization for power: evaluating control-flow-based and event-counter-based techniques," 2006. [Online]. Available: http://parapet.ee.princeton.edu/papers/canturk-hpca2006.pdf. [Accessed 15 8 2019].

[61] O. Khan and S. Kundu, "Microvisor: a runtime architecture for thermal management in chip multiprocessors," 2011. [Online]. Available: https://link.springer.com/chapter/10.1007/978-3-642-24568-8_5. [Accessed 20 11 2019].

[62] S. Srinivasan, R. Kumar and S. Kundu, "Program phase duration prediction and its application to fine-grain power management," 2013. [Online]. Available: https://doi.org/10.1109/isvlsi.2013.6654634. [Accessed 20 11 2019].

[63] N. L. Binkert, B. M. Beckmann, G. Black, S. K. Reinhardt, A. G. Saidi, A. Basu, J. Hestness, D. R. Hower, T. Krishna, S. Sardashti, R. Sen, K. Sewell, M. Shoaib, N. Vaish, M. D. Hill and D. A. Wood, "The gem5 simulator," ACM Sigarch Computer Architecture News, vol. 39, no. 2, pp. 1-7, 2011.

[64] J. Lowe-Power, A. M. Ahmad, A. Akram, M. Alian, R. Amslinger, M. reozzi, A. Armejach, N. Asmussen, B. Beckmann, S. Bharadwaj, G. Black, G. Bloom, B. R. Bruce, D. R. Carvalho and Castrillon, "The gem5 Simulator: Version 20.0+," Arxiv, 2020.

[65] T. Krishna and S. Bharadwaj, "Interconnect Modeling for Homogeneous and Heterogeneous Multiprocessors," in Network-on-Chip Security and Privacy, 2021, pp. 31-54.

[66] J. L. Greathouse and G. H. Loh, "Machine learning for performance and power modeling of heterogeneous systems," In Proceedings of the International Conference on Computer-Aided Design (ICCAD '18). ACM, New York, NY, USA, Article 47, 6 pages, 2018.

[67] S. Bharadwaj, J. Yin, B. Beckmann and T. Krishna, "Kite: A Family of Heterogeneous Interposer Topologies Enabled via Accurate Interconnect Modeling," in 57th ACM/IEEE Design Automation Conference (DAC), San Francisco, CA, US, 2020.

[68] I. Karlin, J. Keasler and J. R. Neely, "LULESH 2.0 Updates and Changes," 2013. [Online]. Available: https://codesign.llnl.gov/pdfs/lulesh2.0_changes.pdf. [Accessed 2 8 2019].

[69] J. Alsop, M. D. Sinclair, S. Bharadwaj, A. Dutu, A. Gutierrez, O. Kayiran, M. LeBeane, B. Potter, S. Puthoor, X. Zhang, T. T. Yeh and B. M. Beckmann, "Optimizing GPU Cache Policies for MI Workloads," in 2019 IEEE International Symposium on Workload Characterization (IISWC), 2019.

[70] BAIDU Research, "DeepBench," [Online]. Available: https://github.com/baidu-research/DeepBench.

[71] S. Dong and D. R. Kaeli, "DNNMark: A Deep Neural Network Benchmark Suite for GPUs," 2017. [Online]. Available: https://dl.acm.org/citation.cfm?id=3038239. [Accessed 2 8 2019].

[72] K. Choi, R. Soma and M. Pedram, "Fine-grained dynamic voltage and frequency scaling for precise energy and performance trade-off based on the ratio of off-chip access to on-chip computation times," 2004. [Online]. Available: http://sportlab.usc.edu/~kihwan/fg-dvfs.pdf. [Accessed 5 7 2019].

[73] V. Spiliopoulos, A. Bagdia, A. Hansson, P. Aldworth and S. Kaxiras, "Introducing DVFS-Management in a Full-System Simulator," 2013. [Online]. Available: http://it.uu.se/katalog/vassp447/gem5_dvfs.pdf. [Accessed 30 7 2019].

[74] Q. Wu, V. J. Reddi, Y. Wu, J. Lee, D. Connors, D. Brooks, M. Martonosi, D. W. Clark and yes, A Dynamic Compilation Framework for Controlling Microprocessor Energy and Performance, 2005, p. 271–282.

[75] Z. Toprak-Deniz, M. A. Sperling, J. F. Bulzacchelli, G. S. Still, R. Kruse, S. Kim, D. W. Boerstler, T. Gloekler, R. Robertazzi, K. Stawiasz, T. Diemoz, G. English, D. T. Hui, P. H. Muench and J. Friedrich, "5.2 Distributed system of digitally controlled microregulators enabling per-core DVFS for the POWER8 TM microprocessor," 2014. [Online]. Available:

http://ieeexplore.ieee.org/document/6757354. [Accessed 25 11 2019].

[76] H. Li, J. Xu, Z. Wang, R. K. V. Maeda, P. Yang and Z. Tian, "Workload-Aware Adaptive Power Delivery System Management for Many-Core Processors," IEEE Transactions on Computer-Aided Design of Integrated Circuits and Systems, vol. 37, no. 10, pp. 2076-2086, 2018.

[77] A. Jog, O. Kayiran, A. Pattnaik, M. Kandemir, O. Mutlu, R. Iyer and C. R. Das, "Exploiting Core Criticality for Enhanced GPU Performance," Sigmetrics Performance Evaluation Review, vol. 44, no. 1, pp. 351-363, 2016.

[78] AMD, "Polaris Whitepaper".

[79] L. Wang, M. Jahre, A. Adileho and L. Eeckhout, "MDM: The GPU Memory Divergence Model," in International Symposium on Microarchitecture(MICRO), 2020.